\documentstyle[12pt]{article}

\begin{document}

\renewcommand{\thefootnote}{\fnsymbol{footnote}}
\renewcommand{\theequation}{\thesection.\arabic{equation}}
\newcommand{\reseteqnum}{\setcounter{equation}{0}}

\newcommand{\calD}{{\cal D}}
\newcommand{\eps}{{\epsilon(z)}}
\newcommand{\epsB}{{\bar \epsilon(\bar z)}}
\newcommand{\epsP}{{\epsilon_+(z)}}
\newcommand{\epsM}{{\epsilon_-(z)}}
\newcommand{\epsPB}{{\bar \epsilon_+(\bar z)}}
\newcommand{\epsMB}{{\bar \epsilon_-(\bar z)}}
\newcommand{\psiL}{{\psi_{\rm L}}}
\newcommand{\psiR}{{\psi_{\rm R}}}
\newcommand{\chiL}{{\chi_{\rm L}}}
\newcommand{\chiR}{{\chi_{\rm R}}}
\newcommand{\rhoL}{{\rho_{\rm L}}}
\newcommand{\rhoR}{{\rho_{\rm R}}}
\newcommand{\tauP}{{\tau_+}}
\newcommand{\tauM}{{\tau_-}}
\newcommand{\Tr}{{{\rm Tr}}}
\newcommand{\VA}{{\rm V / \rm A}}
\newcommand{\nn}{{\nonumber}}
\newcommand{\ra}{{\rightarrow}}

\title{
\hfill
\parbox{4cm}{\normalsize KUNS 1290\\HE(TH)~94/13\\hep-th/9409072}\\
\vspace{2cm}
Axial-Vector Duality as a Mirror Symmetry
\vspace{1.5cm}}
\author{Tomomi Muto\thanks{e-mail address:
\tt muto@gauge.scphys.kyoto-u.ac.jp}\\
{\normalsize\em Department of Physics, Kyoto University}\\
{\normalsize\em Kyoto 606-01, Japan}}
\date{\normalsize September, 1994}
\maketitle
\vspace{1cm}

\begin{abstract}
\normalsize
We study $N=2$ supersymmetric $SU(2)/U(1)$ and $SL(2,R)/U(1)$
gauged Wess-Zumino-Witten models.
It is shown that the vector gauged model is transformed to the
axial gauged model by a mirror transformation.
Therefore the vector gauged model and the axial gauged model
are equivalent as $N=(2,2)$ superconformal field theories.
In the $SL(2,R)/U(1)$ model, it is known that axial-vector duality
relates a background with a singularity to that without a singularity.
Implications of the equivalence of these two models
to space-time singularities are discussed.

\end{abstract}

\newpage

\section{Introduction}

Target-space duality is one of the most striking properties
in string theory \cite{GPR}.
It relates backgrounds with different geometries that are
equivalent as conformal field theories.
This is a characteristic property of string theory,
which is not observed in theories based on point particles.
As in the simplest example, $R \leftrightarrow 1/R$ duality
of a circle compactified model \cite{KYSS},
target space duality generally relates physics
at small scales to that at large scales.
Hence it plays a crucial role in
understanding geometries at short distance, especially issues
of singularities, which is inevitable in general relativity.

{}From this point of view, it is important to study
axial-vector duality in gauged Wess-Zumino-Witten models,
since these models are related to backgrounds with
singularities such as black holes \cite{WittenBH}.
The most interesting example is a $SL(2,R)/U(1)$ model
where axial-vector duality relates a background with a
singularity to that without a singularity.
This is a signal that singularities are smeared out in
string theory.

Recently, another type of duality, mirror symmetry, is studied
mainly in relation to Calabi-Yau manifolds \cite{mirror}.
This is a duality between $N=(2,2)$ superconformal field theories.
In refs. \cite{Henningson,GW}, it is pointed out that
axial-vector duality in $N=2$ supersymmetric gauged
Wess-Zumino-Witten models is nothing but mirror symmetry in
some cases.
This opens a new possibility to study axial-vector duality
from a different standpoint.

In this paper, we present another and more direct method than
refs. \cite{Henningson,GW} to describe the equivalence
between axial-vector duality and mirror symmetry in the
$SU(2)/U(1)$ model.
Then we apply this method to the $SL(2,R)/U(1)$ model.
It is shown that the vector gauged model which corresponds to a
singular background is a mirror partner of the axial
gauged model which corresponds to a non-singular background.
Although this result is already pointed out in ref. \cite{GW},
our method enables us to clarify the reason why the mirror
transformation relates the singular background to the
non-singular background.

We also discuss implications of the result to space-time
singularities.
In fact, it seems that the result leads to a puzzle in the
following sense.
Theories which are related by mirror symmetry should have
the same structures in terms of $N=(2,2)$ superconformal field theories.
Especially, they should have the same chiral ring.
However, in ref. \cite{Eguchi},
it is suggested that the chiral ring is intimately connected with
the existence of space-time singularities in the context of
twisted $N=2$ gauged Wess-Zumino-Witten models.
If this is the case, it seems that a model with a singularity
cannot be a mirror partner of a model without a singularity,
since they might have different chiral ring.
We discuss a possible resolution of this puzzle.

\section{Axial-vector duality and mirror transformation}
\reseteqnum

In this section, let us recapitulate axial-vector duality
and mirror symmetry in the $N=2$ gauged Wess-Zumino-Witten model.
The action of the Wess-Zumino-Witten model \cite{WittenWZ}
on a group manifold $G$ at level $k$ is
\begin{equation}
S(g) = -\frac{k}{8\pi}\int_\Sigma d^2 \sigma\sqrt h h^{ij}
\Tr(g^{-1}\partial_i g\,g^{-1}\partial_j g)-ik\Gamma,
\label{eq:WZW}
\end{equation}
\begin{equation}
\Gamma = \frac{1}{12\pi}\int_{\rm B} d^3 \sigma\epsilon^{ijk}
\Tr(g^{-1}\partial_i g\,g^{-1}\partial_j g\,g^{-1}\partial_k g),
\label{eq:WZ}
\end{equation}
where $\Sigma$ is a Riemann surface, $h_{ij}$ is a metric
on $\Sigma$, $g$ is a map from $\Sigma$ to $G$,
B is any three -manifold whose boundary is $\Sigma$.

Due to the Polyakov-Wiegmann formula \cite{PW}
\begin{equation}
S(gh) = S(g) + S(h) - \frac{k}{2\pi}\int_\Sigma d^2 z
\Tr(g^{-1}\partial_z g\,\partial_{\bar z}h h^{-1}),
\label{eq:PW}
\end{equation}
the action (\ref{eq:WZW}) is invariant under $G_L \times G_R$
transformation, $g \ra v(\bar z)g w(z)$.
Not all of this symmetry can be gauged because of anomaly.
If we want to gauge abelian subgroup $H$, two types of gauging
is allowed \cite{Kiritsis}. One is vector gauging,
\begin{equation}
g \to u^{-1}gu,
\end{equation}
the other is axial gauging,
\begin{equation}
g \ra ugu.
\end{equation}
By introducing gauge fields $A_i$ which transform as
$A_i \ra A_i + u^{-1}\partial_i u$,
the action of the gauged Wess-Zumino-Witten model can be written as
\begin{equation}
S_{\VA}(g,A) = S(g) + \frac{k}{2\pi}\int_\Sigma d^2 z
     \Tr(A_{\bar z}g^{-1}\partial_z g \mp A_z\partial_{\bar z}g g^{-1}
     -A_z A_{\bar z} \pm A_{\bar z}g^{-1}A_z g),
\label{eq:gWZW}
\end{equation}
where upper (lower) sign refer to vector (axial) gauged model.
(The same convention is used in the following.)
We call the relation between these two theories
as axial-vector duality.

The supersymmetric extension \cite{DKPRSchnitzer}
of (\ref{eq:gWZW}) is
\begin{equation}
S_{\VA}(g,\psiL,\psiR,A) =S_{\VA}(g,A)
   +\frac{ik}{2\pi}\int_\Sigma d^2 z \Tr
   \Bigl(\psiL \Bigl(\partial_{\bar z}\psiL +[A_{\bar z},\psiL]\Bigr)
   +\psiR \Bigl(\partial_z \psiR \pm [A_z,\psiR]\Bigr)\Bigr),
\label{eq:sgWZW}
\end{equation}
where $\psiL$ and $\psiR$ are Weyl fermions with values in Lie$(G/H)$.
Here Lie$(G/H)$ represents the orthogonal complement of Lie$H$ in
Lie$G$.
The fermions transform under gauge transformation as
\begin{eqnarray}
\psiL & \ra & u^{-1}\psiL u, \nn \\
\psiR & \ra & u^{\mp 1}\psiR u^{\pm 1}.
\end{eqnarray}

The conditions for the action (\ref{eq:sgWZW}) to have $N=2$
supersymmetry is known \cite{KS,Witten}.
First, Lie($G/H$) must be splitted into two parts,
\begin{equation}
{\rm Lie}(G/H) = \tauP\oplus\tauM,
\label{eq:decomposition}
\end{equation}
where $\tauP$ and $\tauM$ are complex conjugate
representations of $H$.
In addition, the integrability condition,
\begin{equation}
[\tauP,\tauP] \subset \tauP,\qquad
[\tauM,\tauM] \subset \tauM,
\label{eq:integrability}
\end{equation}
and the hermiticity condition,
\begin{equation}
{\rm Tr}(ab)=0, \qquad
{\rm for} \quad a,b \in \tauP \quad {\rm or} \quad a,b \in \tauM.
\label{eq:hermiticity}
\end{equation}
must be satisfied.

We now decompose fermions as $\psiL_{(R)} =\chiL_{(R)} +\rhoL_{(R)}$,
where $\chiL_{(R)}$ and $\rhoL_{(R)}$ belong to $\tauP$ and
$\tauM$ respectively.
By using the equation (\ref{eq:hermiticity}),
the action (\ref{eq:sgWZW}) is written as
\begin{equation}
S_{\VA}(g,\psiL,\psiR,A) =S_{\VA}(g,A)
   +\frac{ik}{2\pi}\int_\Sigma d^2 z \Tr
   \Bigl(\chiL \Bigl(\partial_{\bar z}\rhoL + [A_{\bar z},\rhoL]\Bigr)
   +\chiR \Bigl(\partial_z \rhoR \pm [A_z,\rhoR]\Bigr)\Bigr).
\label{eq:ssgWZW}
\end{equation}
Under the condition (\ref{eq:integrability}),
this action has $N=(2,2)$ supersymmetry \cite{Witten,Nakatsu}.
$N=2$ supersymmetry transformation laws for left moving part are
given by
\begin{eqnarray}
\delta g \; & = & i\epsP g \chiL +i\epsM g \rhoL, \nn \\
\delta\chiL & = & -\frac{i}{k^2} \epsP \chiL\chiL
              +\epsM {\rm P_+}
              \left[g^{-1} (\partial_z g {\pm} A_z g -g A_z)
              -\frac{i}{k^2} (\chiL\rhoL+\rhoL\chiL)\right], \nn \\
\delta\rhoL & = & \epsP {\rm P_-}
              \left[g^{-1} (\partial_z g {\pm} A_z g -g A_z)
              -\frac{i}{k^2} (\chiL\rhoL+\rhoL\chiL)\right]
              -\frac{i}{k^2} \epsM \rhoL\rhoL, \label{eq:susyL} \\
\delta\chiR & = & \delta\rhoR = \delta A_i =0. \nn
\end{eqnarray}
For right moving part,
\begin{eqnarray}
\delta g \; & = & i\epsPB \chiR g +i\epsMB \rhoR g, \nn \\
\delta\chiR & = & \frac{i}{k^2} \epsPB \chiR\chiR
              +\epsMB {\rm P_+} \left[(\partial_{\bar z} g
              {\pm} A_{\bar z} g -g A_{\bar z}) g^{-1}
              +\frac{i}{k^2} (\chiR\rhoR+\rhoR\chiR)\right], \nn \\
\delta\rhoR & = & \epsPB {\rm P_-} \left[(\partial_{\bar z} g
              {\pm} A_{\bar z} g -g A_{\bar z}) g^{-1}
              +\frac{i}{k^2} (\chiR\rhoR+\rhoR\chiR)\right]
              +\frac{i}{k^2} \epsMB \rhoR\rhoR, \label{eq:susyR} \\
\delta\chiL & = & \delta\rhoL = \delta A_i =0. \nn
\end{eqnarray}
Here ${\rm P}_{\pm}$ denote projections to $\tau_{\pm}$.

The action (\ref{eq:ssgWZW}) also has $U(1)$ $R$ symmetry,
whose charge is assigned 1 for $\chiL_{(R)}$, $-1$ for $\rhoL_{(R)}$,
0 for $g$ and $A_i$. Note that $N=2$ supersymmetry transformations
consist of $\Delta R =1$ sector and $\Delta R =-1$ sector.

Let $G_{\pm}(z)$ be the currents associated with the supersymmetry
transformations whose parameters are $\epsilon_{\pm}(z)$,
and $J(z)$ be the $U(1)$ current for left $R$ symmetry.
Then the energy-momentum tensor $T(z)$, the supercurrents
$G_+(z)$, $G_-(z)$ and $U(1)$ current $J(z)$ generate a
$N=2$ superconformal algebra.
Together with a right moving part, the theory has $N=(2,2)$
superconformal symmetry.

The mirror duality is an isomorphism between two $N=(2,2)$
superconformal field theories which differ only in the sign
of the right moving $U(1)$ current \cite{LVW}.
This is a trivial relation from a CFT point of view,
so it is a symmetry between the two theories.
However, target space geometries corresponding to these theories
are quite different in general, since the geometrical meaning
of the operators changes if the relative sign
of right $U(1)$ charge to left $U(1)$ charge is flipped.
The operator product expansions, especially,
\begin{equation}
J(z)G_{\pm}(w) \sim \pm \frac{G_{\pm}(w)}{z-w}
\end{equation}
show that the change in the sign of the $U(1)$ current
corresponds to the interchange between $G_+$ and $G_-$.
It corresponds to a complex conjugation
$\tau_+ \leftrightarrow \tau_-$.
Therefore the mirror transformation is accomplished by taking a
complex conjugation only for right moving part.

\section{The $SU(2)/U(1)$ model}
\reseteqnum

We now show that the vector gauged model is transformed to the axial
gauged model by a mirror transformation in the $SU(2)/U(1)$ model.
The generators of the $SU(2)$ Lie algebra are
\begin{equation}
T=\{T_1,T_2,T_3\}, \qquad T_i=\sigma_i /2,
\end{equation}
where $\sigma_i$ are Pauli matrices.
If we gauge $T_3$ direction, Lie$(G/H)$ is spanned by $\{T_1,T_2\}$.
We can take an almost complex structure $h$ acting on Lie$(G/H)$
as follows,
\begin{equation}
h \, T_1=-T_2, \qquad h \, T_2=T_1.
\end{equation}
With respect to $h$, Lie$(G/H)$ is splitted
into holomorphic and anti-holomorphic parts whose generators are
$T_+ = T_1 +i T_2$ and $T_- = T_1 -i T_2$ respectively.
Then the integrability condition (\ref{eq:integrability})
and the hermiticity condition (\ref{eq:hermiticity}) are
satisfied.
In this case, complex conjugation is realized by a following
operation,
\begin{equation}
T_{\pm} \ra s T_{\pm} s^{-1} = T_{\mp}, \qquad {\rm with} \quad
s=e^{i\pi T_1} = i\sigma_1.
\end{equation}

The partition function of the $SU(2)/U(1)$ vector gauged
Wess-Zumino-Witten model is
\begin{equation}
Z = \int \calD g\,\calD \psiL\,\calD\psiR\,\calD A\,
    {\rm exp}[-S_{\rm V}(g,\psiL,\psiR,A)].
\label{eq:Z}
\end{equation}
As discussed in section 2, the mirror transformation is to take a
complex conjugation only for the right moving part, so the
mirror transformation of the fermions are given by
\begin{eqnarray}
\psiL & \ra & \tilde\psiL = \psiL, \nn\\
\psiR & \ra & \tilde\psiR = s \psiR s^{-1}.
\end{eqnarray}
{}From the requirement of the supersymmetry,
the mirror transformation of the field $g$ is determined as
\begin{equation}
g \ra \tilde g = sg.
\end{equation}
The gauge fields $A_i$ are invariant under the mirror transformation.

Thus the vector gauged model is transformed to
\begin{equation}
Z_{\rm m} = \int\calD \tilde g\,\calD\tilde\psiL\,\calD\tilde\psiR\,
\calD A \, {\rm exp} \,
[-S_{\rm V}(\tilde g,\tilde\psiL,\tilde\psiR,A)],
\label{eq:Zmirror}
\end{equation}
by the mirror transformation.
The classical action is invariant under the vector gauge
transformation,
\begin{eqnarray}
\tilde g \; & \ra & u^{-1} \tilde g u, \nn \\
\tilde \psiL & \ra & u^{-1} \tilde\psiL u, \nn \\
\tilde \psiR & \ra & u^{-1} \tilde\psiR u, \label{eq:gaugemirror}\\
A_i & \ra & A_i +u^{-1} \partial_i u, \nn
\end{eqnarray}
with $u = {\rm exp}(i\epsilon T_3)$.

We now change the variables from
$\tilde g, \tilde\psiL, \tilde\psiR$ to $g, \psiL, \psiR$.
By using the relation $s A_i s^{-1} = -A_i$, which follows from
the fact that the gauge fields are proportional to $T_3$,
we can show that
\begin{equation}
S_{\rm V}(\tilde g,\tilde\psiL,\tilde\psiR,A) =
    S_{\rm A} (g,\psiL,\psiR,A)
    -\frac{k}{2\pi}\int d^2 z \Tr(i \pi T_1 F_{z \bar z}).
\end{equation}
Here $F_{z \bar z} =\partial_z A_{\bar z} -\partial_{\bar z} A_z$
is also proportional to $T_3$.
The last term, which comes from the ``classical'' anomaly of the
bosonic part of the action, vanishes since $\Tr(T_1 T_3) =0$.
Therefore, after change of variables, the classical action of the
mirror transformed model (\ref{eq:Zmirror}) becomes that of
the axial gauged model.

Corresponding to this, the gauge transformation laws for the
fields $g,\psiL,\psiR$ can be read off from (\ref{eq:gaugemirror}).
The result is the axial gauge transformation,
\begin{eqnarray}
g \; & \ra & u g u, \nn \\
\psiL & \ra & u^{-1} \psiL u, \\
\psiR & \ra & u \psiR u^{-1}. \nn
\end{eqnarray}
Here we have used the relation $s u s^{-1} = u^{-1}$.

The Jacobian associated with the change of variables from
$\tilde\psiL, \tilde\psiR$ to $\psiL, \psiR$ can be calculated
by the chiral anomaly,
\begin{equation}
\calD\tilde\psiL \, \calD\tilde\psiR =\calD\psiL \, \calD\psiR \,
{\rm exp} \left[-\frac{1}{4\pi} \int d^2 z \, {\rm Tr} \,
(i \pi T_1 F_{z \bar z})\right].
\end{equation}
Here ``Tr'' represents the trace in the adjoint representation of $G$.
Again, the integrand vanishes since $F_{z \bar z}$ is
proportional to $T_3$.
As for $\calD g$, we can take a Haar measure,
so the integration measures are invariant under change of
variables.

Thus we have shown that the vector gauged model (\ref{eq:Z})
is transformed to the axial gauged model,
\begin{equation}
Z_{\rm m} = \int \calD g \, \calD \psiL \,
\calD \psiR \, \calD A \,
{\rm exp}[-S_{\rm A}(g,\psiL,\psiR,A)],
\end{equation}
by the mirror transformation.
Therefore the $SU(2)/U(1)$ vector gauged model is equivalent to
the $SU(2)/U(1)$ axial gauged model as $N=(2,2)$ superconformal
field theories.

\section{The $SL(2,R)/U(1)$ model}
\reseteqnum

Next, we apply the method explained in the last section to
the $SL(2,R)/U(1)$ case, which is more
interesting physically since it corresponds to a black hole
\cite{WittenBH}.
The generators of the $SL(2,R)$ Lie algebra are
\begin{equation}
T = \{ T_1,T_2,T_3 \}, \qquad T_1={\sigma_1 \over 2}, \;
T_2={i \sigma_2 \over 2}, \; T_3={\sigma_3 \over 2}.
\end{equation}

If we gauge the $T_3$ direction, it is known that the vector
gauged model and the axial gauged model are self-dual, and
correspond to a Lorentzian black hole \cite{GiveonDVV}.
In this case, we cannot take a complex structure which satisfies
the hermiticity condition (\ref{eq:hermiticity}).
Instead, we take an almost product structure.
The possibility of a theory with an almost product
structure to have $N=2$ supersymmetry is discussed in ref.
\cite{GHR}.
An almost product structure ${\sl \Pi}$ is a tensor that maps
Lie($G/H$) to Lie($G/H$) and satisfies ${\sl \Pi}^2 = 1$.
Since the eigenvalue of ${\sl \Pi}$ is $\pm 1$,
Lie($G/H$) is decomposed into two parts $\tau_{\pm}$
according to their eigenvalues.
In the present case, we take an almost product structure
as follows,
\begin{equation}
{\sl \Pi} \, T_1 = T_2, \qquad {\sl \Pi} \, T_2 =T_1.
\end{equation}
The generators of $\tau_{\pm}$ are $T_{\pm} = T_2 \pm T_1$.
It satisfies the integrability condition and the hermiticity
condition which are defined in the same way as
(\ref{eq:integrability}) and (\ref{eq:hermiticity}).
This makes the theory to have $N=2$ supersymmetry.
The discussion about the mirror transformation is
the same as in the $SU(2)/U(1)$ case with a matrix $s$
be replaced by $s ={\rm exp}(\pi T_2) =i\sigma_2$.
In this case, the mirror partners are self-dual to each other,
so the equivalence of the two models under mirror duality is
manifest.

Next we consider the most interesting case, that is, the case of
gauging the $T_2$ direction.
The corresponding background is ``trumpet'' for the vector gauged
model, and ``cigar'' for the axial gauged model \cite{GiveonDVV}.
These backgrounds are called Euclidean black holes.
As we will see later, the ``trumpet'' background has a singularity,
while the ``cigar'' background has no singularity.

In this case, we take a complex structure acting on Lie$(G/H)$ as
\begin{equation}
h \, T_1 = - T_3, \qquad h \, T_3 = T_1,
\end{equation}
then  Lie$(G/H)$ is splitted into $\tau_{\pm}$ whose generators are
$T_{\pm} = T_1 \pm i T_3$.
Complex conjugation is realized as follows,
\begin{equation}
T_{\pm} \ra s T_{\pm} s^{-1} = T_{\mp}, \qquad {\rm with} \quad
s=i e^{-i \pi T_1} =\sigma_1.
\end{equation}

The mirror transformations of $SL(2,R)$ valued field $g$ and
Lie($SL(2,R)/U(1)$) valued fields $\psiL$ and $\psiR$ are
given by
\begin{eqnarray}
g \;  & \ra & \tilde g = sg =\sigma_1 g, \nn \\
\psiL & \ra & \tilde\psiL =\psiL, \\
\psiR & \ra & \tilde\psiR =s \psiR s^{-1}
              =e^{-i \pi T_1} \psiR e^{i \pi T_1}. \nn
\end{eqnarray}
There is a subtlety in this case. That is, the determinant of
$\tilde g$ is $-1$, so $\tilde g$ no longer belongs to $SL(2,R)$
\footnote{This is not because the choice of $s$ is not appropriate.
Although other choices of $s$ are possible to realize the mirror
transformation, the field $\tilde g =sg$ does not belong to
$SL(2,R)$ anyway.}.
So we must carefully analyze the mirror transformation.

We first consider the classical action.
We start from the action of the vector gauged Wess-Zumino-Witten
model,
\begin{equation}
S_{\rm V}(g,\psiL,\psiR,A).
\end{equation}
After the mirror transformation, we obtain the action
\begin{equation}
S_{\rm V}(\tilde g,\tilde\psiL,\tilde\psiR,A).
\label{eq:SmirrorV}
\end{equation}
Here $\tilde g =sg =\sigma_1 g$ is not $SL(2,R)$ valued,
so the action (\ref{eq:SmirrorV}) is not well-defined
as a $SL(2,R)/U(1)$ {\sl vector} gauged Wess-Zumino-Witten model.
However, this is not an assertion that the action
(\ref{eq:SmirrorV}) is not well-defined in any sense.
In fact, after change of variables from
$\tilde g, \tilde\psiL, \tilde\psiR$ to $g, \psiL, \psiR$,
the action is written as
\begin{equation}
S_{\rm A}(g,\psiL,\psiR,A),
\label{eq:SmirrorA}
\end{equation}
as in the $SU(2)/U(1)$ case.
Here $g$ is $SL(2,R)$ valued, so this action is well-defined
as a $SL(2,R)/U(1)$ {\sl axial} gauged Wess-Zumino-Witten model.
Hence, the action (\ref{eq:SmirrorV}), which is related to
(\ref{eq:SmirrorA}) by change of variables, is also
well-defined in this sense.

We next consider the integration measures.
The Jacobian associated with the change of variables from
$\tilde\psiL$, $\tilde\psiR$ to $\psiL$, $\psiR$ can be calculated
as in the $SU(2)/U(1)$ case, and we can see that
$\calD\tilde\psiL\,\calD\tilde\psiR =\calD\psiL\,\calD\psiR$.
As for $\calD \tilde g$, we can not take a Haar
measure for $SL(2,R)$ since det $\tilde g =-1$.
Instead, we take a Haar measure for $GL(2,R)$.
In the parameterization $\tilde g =(\tilde g_{ij}) \in GL(2,R)$,
the Haar measure for $GL(2,R)$ is given by
\begin{equation}
d \tilde g = |{\rm det} \tilde g|^{-2}
            d \tilde g_{11} d \tilde g_{12}
            d \tilde g_{21} d \tilde g_{22}.
\end{equation}
We are now considering the case of det $\tilde g =-1$,
so $d \tilde g = d \tilde g_{11} d \tilde g_{12}
d \tilde g_{21} d \tilde g_{22}$. After change of
variables from $\tilde g$ to $g =s^{-1} \tilde g =\sigma_1 \tilde g$,
it is written as $d \tilde g = d g_{11} d g_{12} d g_{21} d g_{22}$,
which is nothing but the Haar measure for $SL(2,R)$.

Thus, by the change of variables, the mirror transformed model
\begin{equation}
Z_{\rm m} = \int\calD \tilde g\,\calD\tilde\psiL\,\calD\tilde\psiR\,
         \calD A \, {\rm exp} \,
         [-S_{\rm V}(\tilde g,\tilde\psiL,\tilde\psiR,A)],
\end{equation}
is rewritten as the $SL(2,R)/U(1)$ axial gauged
Wess-Zumino-Witten model,
\begin{equation}
Z_{\rm m} = \int\calD g\,\calD\psiL\,\calD\psiR\,
            \calD A \, {\rm exp} \,
            [-S_{\rm A}(g,\psiL,\psiR,A)].
\end{equation}
Therefore the the vector gauged model which corresponds to singular
``trumpet'' background and the axial gauged model which corresponds
to non-singular ``cigar'' background are mirror partner,
and hence they are equivalent as $N=(2,2)$ superconformal field
theories.

We now see the reason why the mirror duality relates the
background with a singularity to that without a singularity.
Space-time singularities in gauged Wess-Zumino-Witten
model occur at fixed points of the gauge transformation.
We parameterize a $SL(2,R)$ valued field $g$ as
\begin{equation}
g = e^{\frac{i}{2}\theta_{\rm L} \sigma_2} e^{\frac{1}{2}r \sigma_1}
    e^{\frac{i}{2}\theta_{\rm R} \sigma_2},
\label{eq:parametrization}
\end{equation}
with $0 \leq r < \infty$, $0 \leq \theta_{\rm L} < 2\pi$,
$-2\pi \leq \theta_{\rm R} < 2\pi$.
Then the vector gauge transformation $g \ra u^{-1} g u$ with
$u ={\rm exp}(i \epsilon \sigma_2/2)$ of the original model
has a fixed point at $r =0$. On the other hand, the {\sl vector}
gauge transformation $\tilde g \ra u^{-1} \tilde g u$ of the
mirror transformed model has no fixed point.
This originates from the peculiar property of the mirror
transformation, that is, det $\tilde g=-1$.

Note that a similar situation takes place in bosonic
three-dimensional black holes \cite{HW}.
In that case, a black string \cite{HH} based on
$SL(2,R) \times U(1)/U(1)$ gauged Wess-Zumino-Witten model,
which has a singularity, is dual to a black hole
without a curvature singularity \cite{BTZ}.
However, the dual model corresponds to an orbifold of a
$SL(2,R)$ Wess-Zumino-Witten model with the $SL(2,R)$ valued
field replaced by a field with negative determinant.
It is interesting to investigate the relation of this duality to
mirror symmetry.

\section{Discussion}
\reseteqnum

We have shown that the vector gauged model corresponding to a
singular background and the axial gauged model corresponding to
a non-singular background are mirror partner.
We now discuss the structure of the singularity which is
expected from the equivalence of the two models under mirror
duality.

In ref. \cite{Eguchi}, it is shown that the singularity
in $SL(2,R)/U(1)$ Lorentzian black hole is described by a
topological field theory and has a structure of chiral ring.
This is based on a twisted version of the $N=2$ gauged
Wess-Zumino-Witten model \cite{Witten}.
By the twisting procedure \cite{WittenEY},
one of the supercharges $G_{\pm}$ becomes a BRS operator
which generates a topological symmetry.
The chiral ring of the model is calculated by a three point
function of BRS invariant operators.
The crucial observation is that the only contribution to the path
integral comes from a fixed point of the BRS transformation,
and furthermore it coincides to the space-time singularity.
It is also speculated that BRS fixed points may generally
occur at space-time singularities.

We now investigate whether this is indeed the case for the
$SL(2,R)/U(1)$ Euclidean black holes.
We twist the vector gauged model by
$T(z) \ra T(z) -\partial_z J(z)/2$, $\bar T(\bar z)
\ra \bar T(\bar z) +\partial_{\bar z} \bar J(\bar z)/2$.
By this procedure, BRS operator is defined as
$Q_B =G_+ +\bar G_-$. The BRS transformations of the fields
can be read off from (\ref{eq:susyL}) and (\ref{eq:susyR}).
By using the fact that $\chiL_{(R)}$ is proportional to
$\sigma_1 +i\sigma_3$, and $\rhoL_{(R)}$ is proportional to
$\sigma_1 -i\sigma_3$, the transformation laws are written as
\footnote{Here we use the same notations for
the fields as in the untwisted model although their conformal
weights are shifted by twisting.}
\begin{eqnarray}
\delta_B g \;  & = & i \epsilon (g \chiL +\rhoR g), \nn \\
\delta_B \rhoL & = & \epsilon P_-
               [g^{-1} (\partial_z g +A_z g -g A_z)], \nn \\
\delta_B \chiR & = & \epsilon P_+
      [(\partial_{\bar z} g +A_{\bar z} g -g A_{\bar z}) g^{-1}], \\
\delta_B \chiL & = & \delta_B \rhoR = \delta A_i =0. \nn
\end{eqnarray}
In the parameterization
\begin{equation}
g =
\left(
\begin{array}{cc}
a  & u \\
-v & b
\end{array}
\right), \qquad {\rm with} \quad ab+uv=1,
\end{equation}
the BRS fixed points for bosonic variables are given by the
following conditions
\begin{equation}
P \partial_z R - R (\partial_z P +i a_z P) =0, \qquad {\rm and} \qquad
Q \partial_{\bar z} R - R (\partial_{\bar z} Q -i a_{\bar z} Q)=0,
\label{eq:BRSFPv}
\end{equation}
with $P=a-b-i(u-v)$, $Q=a-b+i(u-v)$, $R=a+b-i(u+v)$.
Here we put $A_i =a_i T_2$.
There are several branches of possible solutions.
The first branch is $P=Q=0$. In the parameterization
(\ref{eq:parametrization}), $P$ and $Q$ are written as
\begin{equation}
P =-2i \, {\rm sinh} \frac{r}{2} \,
               e^{ \frac{i}{2}(\theta_L-\theta_R)}, \qquad
Q = 2i \, {\rm sinh} \frac{r}{2} \,
               e^{-\frac{i}{2}(\theta_L-\theta_R)}.
\end{equation}
Therefore the solution of this branch is $r=0$.
The second branch of possible solutions of (\ref{eq:BRSFPv})
is $R=0$. However, there is no solution in this branch since
\begin{equation}
R =2 \, {\rm cosh} \frac{r}{2} \,
               e^{-\frac{i}{2}(\theta_L+\theta_R)}
\end{equation}
never vanishes.
Thus the BRS fixed point occurs at $r=0$, just at which a
singularity lies.

Next we study BRS fixed points of the axial gauged model.
Since it is a mirror partner of the vector gauged model,
we must twist as
$T(z) \ra T(z) -\partial_z J(z)/2$, $\bar T(\bar z)
\ra \bar T(\bar z) -\partial_{\bar z} \bar J(\bar z)/2$
in order to compare the results with that of the vector
gauged model twisted as in the last paragraph.
In this case, the BRS operator is defined as
$Q_B =G_+ +\bar G_+$. Then the BRS transformation laws are
\begin{eqnarray}
\delta_B g \;  & = & i \epsilon (g \chiL +\chiR g), \nn \\
\delta_B \rhoL & = & \epsilon P_-
               [g^{-1} (\partial_z g -A_z g -g A_z)], \nn \\
\delta_B \rhoR & = & \epsilon P_-
      [(\partial_{\bar z} g -A_{\bar z} g -g A_{\bar z}) g^{-1}], \\
\delta_B \chiL & = & \delta_B \chiR = \delta A_i =0. \nn
\end{eqnarray}
The BRS fixed points are given by
\begin{equation}
R \partial_z P - P (\partial_z R +i a_z R)= 0, \qquad {\rm and} \qquad
S \partial_{\bar z} P - P (\partial_{\bar z} S -i a_{\bar z} S) =0,
\label{eq:BRSFPa}
\end{equation}
with $S=a+b+i(u+v)$.
The first branch of possible solutions is $R=S=0$.
However, there is no solution in this branch since $R$ and
\begin{equation}
S =2 \, {\rm cosh} \frac{r}{2} \,
               e^{\frac{i}{2}(\theta_L+\theta_R)}
\end{equation}
never vanish. The second branch is $P=0$, and it corresponds to
$r=0$. However, this branch does not contribute to the
calculation of chiral ring by the following reason.
We can see that $P$ is a BRS invariant operator by analyzing
$\delta_B g$.
Since the chiral ring is calculated by a three point function
of BRS invariant operators, the integrand vanishes just at
which the contribution to the path integral comes.

These results are consistent with the speculation that BRS fixed
points may occur at space-time singularities.
However, it seems that this leads to an asymmetry in the
chiral ring between mirror partners.
This is inconsistent with the result that the singular
``trumpet'' background and the non-singular ``cigar'' background
are mirror partners, and hence equivalent as $N=(2,2)$ superconformal
field theories. This is a puzzle I mentioned in introduction.

Exactly speaking, however, there is another possibility to
solve the equations (\ref{eq:BRSFPv}) or (\ref{eq:BRSFPa}).
For the vector gauged model, it is given by the following equations,
\begin{equation}
P=fR \quad {\rm with} \quad \partial_z f +i a_z f=0, \qquad
           {\rm and} \qquad Q=hR \quad {\rm with}
           \quad \partial_{\bar z} h -i a_{\bar z} h =0.
\end{equation}
We note that this branch does not consist of classical solutions
of the gauged Wess-Zumino-Witten model, so the contributions
would vanish exponentially for $k \ra \infty$ \cite{Witten}.
A similar possibility is also allowed for the axial gauged model.
The contribution from this branch, which can be interpreted
as a quantum correction to the structure of the singularity,
may compensate for the apparent asymmetry in the chiral ring
between mirror partners.

Note that a similar situation occurs in a discussion of
topology change of Calabi-Yau manifolds \cite{AGMWitten}.
In the course of topology change,
target manifolds necessarily go through a singular configuration,
while the mirror partner does not. In that case,
non-perturbative world-sheet instanton corrections \cite{DSWW}
smooth a discreteness associated with the singularity,
and mirror symmetry is preserved.

To make a definite statement, we must further investigate a three point
function of BRS invariant operators, especially the contribution
from the third branch.

\vskip 1cm
\centerline{\large\bf Acknowledgements}

I would like to thank K.~Sakai, S.~Yahikozawa and Y.~Yoshida
for valuable discussions.

\newcommand{\J}[4]{{\sl #1} {\bf #2}(19#3) #4}
\newcommand{\NP}{Nucl.~Phys.}
\newcommand{\PL}{Phys.~Lett.}
\newcommand{\PR}{Phys.~Rev.}
\newcommand{\PRL}{Phys.~Rev.~Lett.}
\newcommand{\MPL}{Mod.~Phys.~Lett.}
\newcommand{\PTP}{Prog.~Theor.~Phys.}
\newcommand{\CMP}{Comm.~Math.~Phys.}
\newcommand{\PRep}{Phys.~Rep.}

\end{document}